\documentclass[runningheads]{llncs}
\usepackage{cite}
\usepackage{amsmath,amssymb,amsfonts}
\usepackage{algorithmic}
\usepackage{graphicx}
\usepackage{textcomp}
\usepackage{xcolor}
\usepackage{booktabs}
\usepackage{multirow}
\usepackage{longtable}
\usepackage{hyperref}
\usepackage{todonotes}
\usepackage{threeparttable}
\usepackage{tikz}
\usepackage[table]{xcolor} 
\usetikzlibrary{shapes,arrows.meta,positioning,calc,fit,backgrounds}
\pgfdeclarelayer{background}
\pgfsetlayers{background,main}
\usepackage{adjustbox}
\usepackage{booktabs}
\usepackage{amsmath,amssymb}
\usepackage{tabularx}
\usepackage{listings}
\usepackage{xcolor}

\lstdefinestyle{xcodelight}{
  basicstyle=\ttfamily\footnotesize,
  backgroundcolor=\color{black!3},
  frame=single,
  rulecolor=\color{black!20},
  columns=fullflexible,
  breaklines=true,
  breakatwhitespace=true,
  showstringspaces=false,
  tabsize=2,
  keepspaces=true,
}

\begin{document}

\title{Comparative Analysis of Cross-Chain Token Standards}

\author{Fatemeh Heidari Soureshjani \and
Jan Gorzny
}
\authorrunning{F.~Soureshjani and J.~Gorzny}

\institute{Zircuit\\
\email{\{fatemeh, jan\}@zircuit.com}} 

\maketitle

\begin{abstract}
Cross-chain token standards enable fungible tokens that exist  across multiple blockchains with a unified total supply model.
This paper presents a comprehensive comparative analysis of five leading cross-chain token standards and frameworks: the xERC20 standard (implementing ERC-7281), the Omnichain Fungible Token (OFT) standard, the Native Token Transfers (NTT) framework, the Cross-Chain Token (CCT) standard, and the SuperchainERC20 standard (implementing ERC-7802). 
We examine each standard’s distinguishing properties and technical design, including architecture, message-passing mechanisms, interoperability scope, chain compatibility, and security features.
Our analysis reveals that while all these standards share the goal of seamless cross-chain fungibility, they differ significantly in implementation approach, trust model, and target ecosystem. 
\end{abstract}

\section{Introduction}\label{sec:intro}

The blockchain ecosystem has evolved from a single-chain paradigm to a highly fragmented multi-chain landscape. 
There are now hundreds of active Layer ~ (L1) blockchains, sidechains, and Layer~2 (L2) networks (see, e.g., \cite{DBLP:conf/fc/GudgeonMRMG20}).
Each of these offers different trade-offs in throughput, cost, security, and functionality.
While this diversity enables innovation and scalability, it also creates significant challenges for token projects and users. 

When a token which was originally deployed on e.g., Ethereum, exists on multiple chains, there must be representatives of that token on all of those additional chains.
This is often achieved by ``lock-and-mint'' bridging that introduces representations of the Ethereum-deployed token on alternative networks.
In this setting, a bridge locks the original supply in a contract on the source chain and mints an equivalent amount of the so-called \emph{wrapped token} on a destination chain~\cite{defi-sok-gogol}. 
These types of bridges are common and indeed are the kind that form the majority of L2 canonical bridges.
Ultimately this approach means that there may be multiple copies of a given token in existence at the same time, and that different bridges may use different representations on the destination chains.

This results in token identity confusion and liquidity fragmentation.
A single token may exist as the original token where it was first deployed (which we call the \emph{native} version); a wrapped version on chain $X$ minted via a bridge $A$ and another wrapped version on chain $X$ minted via a bridge $B$; a wrapped version on chain $Y$ minted via bridge $C$; and more.
Now liquidity may also be fragmented: e.g., such a token with \$100M total liquidity might have \$40M on Ethereum, \$25M on Arbitrum, \$20M on Optimism, and \$15M on Polygon.
This proliferation of token representations blurs the notion of which version is ``canonical.''
Furthermore, it complicates integrations for wallets and exchanges, as well as forcing users to understand complex bridging mechanics.

This fragmentation leads to reduced capital efficiency, price discrepancies, degraded user experience, and increased Decentralized Finance (``DeFi''; see, e.g., \cite{defi-sok-gogol}) complexity.
This means that actors like market makers and liquidity providers (LPs) must split their capital among multiple venues, different representations of the same token can have different prices, users are unclear about which tokens are legitimate or equivalent, and onchain protocols must adapt for multiple implementations.
Moreover, each bridge must now account for custodial risk and users of these representations must be aware of the risks introduced by a particular bridge design.

Cross-chain token standards allow a fungible token to exist natively across multiple blockchains and attempt to solve the aforementioned issues. 
Unlike tokens which are first deployed on a single blockchain, and then represented via wrapper tokens on other blockchains, these standards aim to allow a unified supply model without identifying a particular implementation and blockchain as the native representation of the token.
These tokens skirt the requirement for wrapped representations by using official smart contract implementations on all of the blockchains that their issuer chooses to support.
While wrapped representations may still be required when these assets are bridged to yet more blockchains, these standards simplify the token landscape on their officially supported chains.
This simplification is achieved by the issuer maintaining several official implementations, possibly in different smart contract languages.
Then, cross-chain tokens are moved via a ``burn-and-mint'' mechanism where tokens are burned on the source chain and minted on the destination, keeping total supply constant across chains.

The use of several official implementations may increase the overhead for the issuer, but ultimately provides a better user experience.
Cross-chain tokens ensure one canonical token per chain and avoids user-facing swaps or liquidity pools during transfers.
Nonetheless, not all approaches to cross-chain tokens are equal and there is no single approach used by all cross-chain token issuers.
As a result, application builders and users still need to understand this complex token environment.

This paper compares five leading cross-chain token standards and frameworks: the xERC20~\cite{xerc20_site} standard (which implements ERC-7281~\cite{xerc20-standard}), the Omnichain Fungible Token (OFT) standard~\cite{layerzero-oft}, the Native Token Transfers (NTT) framework~\cite{wormhole_ntt} (which  requires a standard token interface to be used), the Cross-Chain Token (CCT) standard~\cite{chainlink-cct}, the SuperchainERC20 standard~\cite{optimism-superchain} (which implements ERC-7802~\cite{eip7802}).
We examine each standard's distinguishing properties and technical design (architecture, message passing, interoperability model, chain compatibility, and security). 
Along the way, we report on usage statistics for these token standards. 
We measure the prevalence of each standard and evaluate how some claims -- like the ability to use permissionless message relayers -- are reflected in practice.
To our knowledge this is the first paper which compares properties of, and reports on the usage of, these tokens.

The remainder of this paper is organized as follows: Section~\ref{sec:related_work} reviews related work. 
Sections~\ref{sec:xerc20}--\ref{sec:cct} each discuss a single standard.
Within each section, we describe the pertinent implementation details, distinguishing properties, and security concerns. 
Section~\ref{sec:comparison-evaluation-discussion} presents a comparative evaluation of the standards relative to each other and comments on their usage, including limitations of the study.
Section~\ref{sec:conclusion} concludes the paper and outlines future research directions.
Reference implementation details are found in Appendix~\ref{sec:impl-details}.

\section{Related Work}
\label{sec:related_work}

Blockchain interoperability has been extensively studied, with numerous surveys analyzing cross-chain communication protocols, bridge architectures, and consensus mechanisms~\cite{ren2023interoperability, 
peelam2024unlocking, 
schulte2019towards, 
belchior2021survey, 
khan2021towards, 
deng2025enhancing, 
lohachab2021towards, 
harris2023cross, 
qasse2019inter, 
yamamoto2023examination
}. 
However, these works focus on protocol-level bridging solutions.
These approaches include atomic swaps, relays, sidechains, and messaging networks, but do not provide focused analyses of cross-chain token standards.

A subset of the literature mentions token formats in interoperability scenarios. 
Belchior et al.~\cite{belchior2021survey} identify ERC-20~\cite{ERC-20} and ERC-721~\cite{ERC-721} as common formats for cross-chain asset transfer but do not examine their extensions as we do in this work. 
Wang et al.~\cite{wang2021sok, wang2023exploring} examine ERC-20 transfer modalities (e.g., ERC-20-to-Native, ERC-20-to-ERC-20) in Proof-of-Authority sidechains. 
Delgado-von-Eitzen et al.~\cite{delgado2025bridging} propose an ``Inter-Blockchain Communication'' relay architecture enabling ERC20-compliant token transfers, but again do not discuss specifics for new cross-chain standards.
Therefore none of these works systematically analyze token standard design or compare emerging standards like in this work; as a result, these prior works are complementary to ours.

\section{xERC20 Tokens (ERC-7281 Tokens)}
\label{sec:xerc20}

xERC20 tokens are an implementation of the ERC-7281 ``Sovereign Bridge Token''~\cite{xerc20-standard} open standard by Connext~\cite{connext_xerc20}.
These tokens extend the ERC-20 interface with controlled burn-and-mint functionality.
Rather than relying on external bridge custody or liquidity pools, xERC20 defines how token supply may be adjusted across chains under issuer-defined constraints. This design shifts interoperability from a bridge-level concern to a token-level abstraction, making trust assumptions explicit and configurable by the token issuer.

\subsection{Mechanism}

xERC20 extends the base ERC-20 interface by adding controlled mint and burn capabilities that authorized bridges can invoke. 
The key characteristic of xERC20 is that it adapts commonly implemented minting and burning functionality in ERC-20 tokens\footnote{Not technically required; see Appendix~\ref{sec:erc20tokens}.} enable controlled and secure cross-chain compatibility.
The \texttt{mint} and \texttt{burn} functions are modified so that they can only be invoked by whitelisted bridge contracts.
Each authorized bridge is not merely whitelisted but is also assigned a specific minting and burning capacity determined by the token issuer.
Therefore xERC20 token must implement the standard ERC-20 interface together with the functions \texttt{setLimits}, \texttt{mintingMaxLimitOf}, \texttt{mintingCurrent\-LimitOf} (and corresponding opposites for burning functionality). 
In xERC20, supply management remains inside the token contract itself.
Once a burn–mint action completes, the bridge has no continuing custody or control over the asset.

Each single-chain ERC-20 token (a.k.a a \emph{legacy token}) has a chain on which it is considered native; any representation on other chains created through traditional bridging is effectively a wrapped version. 
To transform the token into one implementing the xERC20 standard, xERC20 introduces a \texttt{lockbox} mechanism to upgrade existing ERC-20 deployments. 
The legacy token is deposited into the \texttt{lockbox} contract on its native chain, and a new xERC20 contract is deployed there with a reference to this lockbox. 
Additional xERC20 contracts are then deployed on the non-native chains. 
Once this migration path is established, the xERC20 representation becomes fungible across all participating chains.
However, this effectively re-introduces ``lock-and-mint'' functionality. 

\subsection{Security Considerations}
\label{sec:xerc20security}

Bridging entry points are permissioned: only whitelisted bridge contracts may call them, and each bridge is assigned a rate-limited minting and burning capacity by the token issuer. 
This allows the token issuer to restrict the influence of any interoperability provider while maintaining a unified supply across chains.

Each authorized bridge is not merely whitelisted but is also assigned a specific minting and burning capacity determined by the token issuer.
These per-bridge rate limits are configured through the \texttt{setLimits} function, which defines how much a given bridge is permitted to mint or burn over time. 
These capacities are upgradeable by the token issuer.

\section{Omnichain Fungible Tokens (OFT)}
\label{sec:oft}

The Omnichain Fungible Token (OFT) is a cross-chain token framework introduced by LayerZero that embeds multi-chain transfer functionality directly into the token contract~\cite{layerzero-oft}. 
Unlike token-centric standards that remain protocol-agnostic, OFT is tightly coupled to the LayerZero messaging stack and derives its security and execution guarantees from LayerZero's verification infrastructure. 

\subsection{Mechanism}
The OFT standard also extends ERC-20 token with burn-and-mint supply synchronization while embedding LayerZero’s multi-chain messaging directly into the token contract. 
Critically, an OFT contract is simultaneously an ERC-20 token and a LayerZero \texttt{OApp}, meaning that the token itself can send and receive cross-chain messages through LayerZero endpoints.

To understand the OFT workflow, it is useful to recall the basics of LayerZero’s multi-chain messaging protocol. 
Each supported chain deploys a LayerZero endpoint contract exposing standardized send and receive entrypoints. 
\texttt{OApp}-based contracts interact with these endpoints to
communicate across chains. 

An OFT contract inherits functionality from the \texttt{OFTCore} interface, which in turn inherits from the \texttt{OApp} interface.
Collectively, these interfaces implement that the token logic required for multi-chain operation: OFT message construction, cross-chain transfer encoding, decimal normalization, dust\footnote{Small token amounts of value below the cost required to transfer it.} removal, and the abstract \texttt{\_debit} (for burning on the source chain) and \texttt{\_credit} (for minting on the destination chain) hooks that define how balances are adjusted on each chain during a transfer.
Because OFT contracts inherit from \texttt{OApp}, they participate in this messaging flow and can initiate and receive multi-chain transfers natively.

The user-facing entry point to an OFT contract is the public \texttt{send()} function.
Unlike bridge-specific routers, OFT exposes a direct interface in which the user (\texttt{msg.sender}) specifies all parameters of the cross-chain transfer, including the destination chain ID, the destination address that will receive the tokens. 
The OFT contract then wraps these parameters into an \texttt{OApp} message and dispatches it through the LayerZero endpoint. 
Because the user initiates the transfer directly on the OFT contract, the trust boundary is minimized, relative to router-based bridges, though it remains dependent on configured DVNs (see Section~\ref{sec:xoftsecurity}).
The OFT does not infer destinations or amounts from any external registry, and the sender’s intent is encoded explicitly in the
message payload.

Similar to the xERC20 standard, legacy tokens can be transformed into those which implement the OFT interface.
This is done by deploying an \texttt{OFTAdapter} contract, which inherits \texttt{OFTCore}, effectively bridging the original token across chains.
Just as in the case of xERC20 tokens, this reintroduces ``lock-and-mint''-like semantics to bridging; such tokens are released when this chain is the destination chain.
However, the \texttt{OFTAdapter} is only deployed native chain for the legacy tokens. 
Therefore on all non-native chains, a standard \texttt{OFT} instance is deployed instead, which still uses ``burn-and-mint'' bridging semantics.
This design ensures that the native chain never mints synthetic supply: the adapter locks real tokens on the native chain, while remote chains mint and burn synthetic OFT supply according to LayerZero messages. 
As a result, the global token supply remains consistent across all chains.

\subsection{Security Considerations}
\label{sec:xoftsecurity}

Cross-chain messages are verified by Decentralized Verifier Networks (DVNs). 
Token issuers configure security by specifying required DVNs (all must sign) and optional DVNs (a threshold subset must sign).
DVN types include oracle-based (Google Cloud, Chainlink), zero-knowledge proof (Polyhedra), and bridge-based (Axelar). 
This $M$-of-$N$ model allows projects to balance security, cost, and latency while avoiding single-point-of-failure risk.
Typically $N \ge M=2$ and there are more than $30$ DVNs available at the time of writing.

\section{Native Token Transfers (NTT) Framework}
\label{sec:ntt}

NTT refers to Wormhole's Native Token Transfers framework to make any token natively multi-chain using the Wormhole cross-chain messaging protocol~\cite{wormhole_ntt}. 
The framework builds on Wormhole's existing bridging infrastructure but improves it by preserving the token's native contract and characteristics on each chain~\cite{wormhole_quicknode}.
While the NTT framework is not a token standard per se, its use requires multi-chain tokens to conform to specified interfaces as described below.

\subsection{Mechanism}
NTT cross-chain functionality operates in two supply-management modes: \emph{Hub-and-spoke} and \emph{Burn-and-Mint}. 
For existing ERC-20 tokens that cannot modify their supply, NTT uses a \emph{Hub-and-spoke} model: tokens on the native chain are locked in the chain's \texttt{NTTManager}, and an equivalent balance is minted on non-native chains. For newly issued tokens NTT uses a Burn-and-Mint model, in which tokens are burned on the source chain and minted on the destination chain. 
Either way, a token registers to protocol through a \texttt{NTTManager} contract. 
There is one \texttt{NTTManager} contract per-chain, per-token and it is responsible for supply management, transfer limits, and Wormhole message handling. 
During a cross-chain transfer, the source-chain \texttt{NTTManager} locks or burns tokens and emits a Wormhole message; upon verification on the destination chain, the corresponding \texttt{NTTManager} mints or unlocks the appropriate amount.


Regarding token requirements, the NTT mode determines which token interface must be implemented. 
In Burn-and-Mint mode, the token must implement the \texttt{INttToken} interface to expose \texttt{mint} and \texttt{burn} capabilities, and the \texttt{NTTManager} must be granted the minter role.
Such a token contract is deployed on all chains participating in the NTT deployment. 
In Lock-and-Mint mode, any ERC-20 token may serve as the canonical token on its native chain registered in the \texttt{NTTManager}, while all non-native chains implement and deploy a corresponding \texttt{INttToken} representation for which that chain's \texttt{NTTManager} must be granted the minter role.

The \texttt{NTTManager} contract on each chain also maintains a list of peer \texttt{NTTManagers} for the same token on all supported chains (and this must be configured by the token issuer on supported chains).

\subsection{Security Considerations}
\label{sec:nttsecurity}

In bound and outbound limits are simply a specified rate of tokens that can arrive on, or leave from, a chain over a period of time.
However, it is important to note that messages sent that are later blocked due to these limits are placed in a queue and may be canceled or processed later.

Moreover, each \texttt{NTTManager} acts as a \texttt{TransceiverRegistry}, enabling the token owner to configure the required attestation threshold by registering multiple \texttt{Transceiver} contracts. 
A \texttt{Transceiver} contract is responsible for relaying and executing cross-chain messages on the destination chain after receiving an on-chain attestation.
An on-chain attestation comes from the form of a ``Verified Action Approvals'' produced by Wormhole guardians watching the source chain for initiations of cross-chain transfers.
Once a VAA signed by a quorum of Wormhole Guardians is produced, any transceiver may submit it for execution on the destination chain.
By default, attestations from $13$ of $19$ Guardians are required to complete the transfer on the destination chain.

\section{Cross-Chain Tokens (CCT)}
\label{sec:cct}

The Cross-Chain Token (CCT) standard by Chainlink aims to offer a ``streamlined and decentralized'' approach to cross-chain token transfers~\cite{chainlink-cct}.
Meant for use with Chainlink's Cross-Chain Interoperability Protocol (CCIP), supported tokens must implement a set of functions though they are mostly managed by other contracts in the CCIP infrastructure.
Having a brief understanding of CCIP is necessary to understand CCT tokens used within it. 
At a high level, CCIP is organized around a \texttt{Router} contract, which acts as the generic entry point for all cross-chain requests. 
Every user-initiated CCIP transaction, including a token transfer or an arbitrary message, is first submitted to the \texttt{Router}. 
The Router is responsible for delegating the request to the appropriate \texttt{OnRamp} or \texttt{OffRamp} contract; in turn, they interact with the appropriate \texttt{TokenPool} contract for each CCT on the relevant chain.
On ramp contracts trigger burn or lock operations on a token pool contract and prepares and formats the message for cross-chain delivery, while off ramp contracts validate delivered messages and unlock or mint tokens.

\subsection{Mechanism}

A token becomes compatible with CCT by implementing a small set of functions grouped into two requirement categories: registration functions and transfer functions. 
Registration primarily establish roles for the token contract; namely, an \texttt{owner} and a \texttt{CCIPAdmin}.
Transfer functions define the minimal interface required to support cross-chain supply adjustments. 
The required functions depend on the chosen supply-handling mode and correspond to either the Burn/Mint requirements or the Lock/Release requirements for the protocol.
For example, in the Burn/Mint setting, functions like \texttt{burn}, \texttt{burnFrom}, and \texttt{mint} are required.
These do not deviate from many standard ERC-20 implementations.

On each chain, there is a \texttt{TokenPool} contract that CCIP interacts with during cross-chain transfers. 
To enable this interaction, each \texttt{TokenPool} instance must implement the token-sending functionality on the source chain and the token-receiving functionality on the destination chain. 
In the standard implementation, each \texttt{TokenPool} contract manages a single ERC-20 token, but the architecture allows token issuers to extend the design and support multiple tokens if needed.

The token pools and cross-chain supply management for CCIP using CCT supports multiple modes including \emph{Burn-and-Mint}, \emph{Lock-and-Mint}, \emph{Burn-and-Unlock}, and \emph{Lock-and-Unlock}. 

CCT enforces ``Decentralized Oracle Network'' (DON) consensus on a per-lane basis, where issuer has no configurability power.
Each lane is a path between two specific blockchains which has been setup and is maintained by the CCIP.
Token issuers cannot choose which DONs are used, adjust quorum thresholds, or change security assumptions.
A cross-chain transfer is executed only after multiple DONs independently observe and attest to the same event.

\subsection{Security Considerations}
\label{sec:xcctsecurity}

In addition to the \texttt{TokenPool} contracts, a \texttt{Token\-AdminRegistry} contract is necessary essential to manage tokens registration and stores the token pool configuration for all CCIP enabled tokens.
This requires additional configuration and also provides methods for rate limiting tokens, as for other cross-chain tokens.

\section{SuperchainERC20 Tokens (ERC-7802 Tokens)}
\label{sec:superchain}

SuperchainERC20 tokens are those that use the Optimism implementation of the ERC-7802~\cite{eip7802} \emph{only} within the Optimism Superchain ecosystem~\cite{optimism-interop}.
The Superchain is a network of OP stack-based L2s which share a unified bridging protocol on a common development stack.
It defines a minimal interface for cross-chain minting and burning on OP Stack chains. 
The goal is to remove liquidity fragmentation \emph{inside} the Superchain by providing a standard supply model supported by OP Stack's shared security and built-in messaging infrastructure.

\subsection{Mechanism}
The \texttt{SuperchainERC20} contract implements the \texttt{IERC7802} interface, a minimal standard that exposes two privileged entrypoints, \texttt{crosschainMint} and \texttt{crosschainBurn}. 
These functions allow the authorized (Superchain) bridge contract to adjust token supply during cross-chain transfers.
No messaging or verification logic is implemented within the token itself; the
Superchain bridge is solely responsible for validating withdrawals and invoking the corresponding mint or burn on remote chains.

\texttt{SuperchainTokenBridge} is the contract through which users initiate
cross-chain ERC--20 transfers. A transfer begins when a user calls \texttt{sendERC20}, which immediately invokes \texttt{crosschainBurn} on the
source-chain token, destroying the specified amount. 
The bridge then emits a \texttt{SendERC20} event for the OP Stack cross-chain messaging layer.
Once the message is verified on the destination chain, the bridge calls \texttt{relayERC20}. 
This function completes the transfer by invoking \texttt{crosschainMint} on the destination-chain token contract, minting the corresponding amount for the recipient. 
The \texttt{SuperchainERC20} contract thus remains a minimal mint--burn interface, while the bridge implements all verification and message handling.

Note that the \texttt{SuperchainERC20} tokens do not implement any cross-chain logic themselves; instead, they integrate with the broader OP Stack interoperability framework through the supplied but generic \texttt{L2ToL2CrossDomainMessenger} contract. 
This messenger provides a unified message passing interface for all contracts that perform L2-to-L2 communication within the Superchain. 
When a user calls \texttt{sendERC20}, the \texttt{SuperchainTokenBridge} invokes \texttt{crosschainBurn} on the local token, constructs the transfer message, and forwards it to the \texttt{L2ToL2CrossDomainMessenger}, which is responsible for routing the message to the destination chain. 
In the destination chain, upon receiving a message and some general verifications, the messenger triggers \texttt{relayMessage}, causing the remote \texttt{SuperchainTokenBridge} to call \texttt{crosschainMint} on the destination token contract. 
Thus, the bridge contracts perform only supply adjustments and message preparation, while the messenger implements the Superchain messaging pipeline.

An additional design constraint is that \texttt{SuperchainTokenBridge} does not maintain a token registry or peer-address mapping. 
Instead, the protocol assumes that each SuperchainERC20 deployment has the same deterministically determined address on every OP Stack chain, allowing the bridge to identify the corresponding token instance without storing any per-token configuration. 
This requirement is satisfied automatically for new tokens that adopt the ERC-7802 standard and deploy at deterministic addresses across chains.
For existing tokens that do not share a unified deployment strategy or deterministic addressing, the Superchain does not provide a standard adapter workflow.

\subsection{Security Considerations}
\label{sec:superchainecurity}

SuperchainERC20 tokens do not have any built-in rate limiting or other security features.
The trusted actors involved in these tokens are the deployer and the Superchain infrastructure; there are no additional third-parties unless the token developer adds in esoteric functionality.
Security assumptions are inherited from the OP Stack fault-proof and messaging model.

\section{Comparison, Evaluation, and Discussion}
\label{sec:comparison-evaluation-discussion}

\subsection{Comparison}
\label{sec:comparison}

Table~\ref{tab:xc-token-matrix} summarizes the key differences between the token standards studied.
In each following subsection, we comment on additional observations from the table that were not captured in prior sections.

\begin{table*}[bt]
\centering
\caption{A summary of token standard features and functionality. In this table, ``configurable'' means per-token, rather than per-bridging protocol (e.g., Wormhole may rotate its guardian set, but tokens do not have the power to change the set).}
\label{tab:xc-token-matrix}
\centering
\makebox[0pt]{
\begin{tabular}{l c c c c c}
\hline 
\textbf{Capability} &
\textbf{xERC20} &
\textbf{OFT} &
\textbf{NTT} &
\textbf{CCT} &
\textbf{SuperchainERC20} \\
\hline 
\hline 
\multicolumn{6}{l}{\textbf{Architecture}} \\
Bridge-agnostic design                & \checkmark & $\times$ & $\times$ & $\times$ & $\times$ \\
Unified native supply                 & \checkmark & \checkmark & \checkmark & \checkmark & \checkmark \\
Support for ERC-20 tokens & \checkmark & \checkmark &  \checkmark & \checkmark & \checkmark \\
Support for ERC-1155 tokens & $\times$ & $\times$  & $\times$  & $\times$ & $\times$ \\
Support for ERC-721 tokens & $\times$ & $\times$  & $\times$  & $\times$  & $\times$ \\ \hline

\multicolumn{6}{l}{\textbf{Bridging Semantics}} \\
Burn-and-mint capability          & \checkmark & \checkmark & \checkmark & \checkmark & \checkmark \\
Lock-and-mint capability              & \checkmark     & \checkmark & \checkmark & \checkmark &  $\times$ \\ 
Support for EVM Chains   & $\times$ & \checkmark & \checkmark  & \checkmark & \checkmark$^*$ \\
Support for Non-EVM Chains   & $\times$   & \checkmark & \checkmark &  \checkmark &  $\times$ \\ 
Whitelist/Blacklist of addresses  &  $\times^*$   & $\times^*$  & $\times^*$ &   $\times^*$ & $\times^*$  \\ \hline 
\multicolumn{6}{l}{\textbf{Control \& Governance}} \\
Issuer-owned contracts                & \checkmark & \checkmark & \checkmark & \checkmark & \checkmark \\ 
Upgradeable contracts                & \checkmark & \checkmark & \checkmark & \checkmark & \checkmark \\ 
Configurable parameters (e.g., decimals)              & \checkmark & \checkmark$^*$ & \checkmark$^*$ & \checkmark$^*$ & \checkmark \\ 
\hline

\multicolumn{6}{l}{\textbf{Fees}} \\
No protocol-level fees & \checkmark & $\times$ & \checkmark &   $\times$ & \checkmark \\
Source-chain only fees & \checkmark & $\times$ & \checkmark & $\times$  & \checkmark \\ \hline 

\multicolumn{6}{l}{\textbf{Security Model}} \\
Rate limits        & \checkmark & \checkmark & \checkmark & \checkmark & $\times$ \\ 
Configurable quorum threshold              & \checkmark$^*$     & $\times$  & $\times$ &  $\times$   & $\times$ \\ 
Configurable quorum participants   & \checkmark  & \checkmark  & $\times$ & $\times$  & $\times$ \\ 
Pausable destination contracts   & \checkmark & \checkmark & \checkmark & \checkmark$^*$  & \checkmark \\ \hline
\end{tabular}
}
\end{table*}

\subsubsection{Architecture.}
Unsurprisingly, none of the standards support additional token standards.
While most were designed specifically for ERC-20 tokens, it's interesting to note that none of the designs were abstract enough to support additional popular standards like ERC-721 (non-fungible tokens; NFT)~\cite{ERC-721} or ERC-1155 (multi-type tokens)~\cite{ERC-1155}.
Furthermore, note that xERC20 is the only standard that is bridge agnostic and is therefore arguably the most decentralized.

\subsubsection{Bridging Semantics.}
Most tokens support both Ethereum Virtual Machine (EVM) chains and non-EVM chains. 
Notable exceptions are xERC20 which could be adapted but for which no adaptations are known, and SuperchainERC20s which only support the EVM chains of the Superchain (although the same code could be deployed on other EVM chains).

Additionally, most tokens can build in functionality for address whitelisting or blacklisting and remain compatible with each standard, but no standard (or corresponding protocol) enforces or requires such functionality.

\subsubsection{Control \& Governance.}
All standards use some form of issuer-deployed contract which must conform to a given interface.
That provides great flexibility: these contracts can technically be upgradeable (behind a so-called proxy contract~\cite{DBLP:journals/ese/EbrahimiAOH24}), pausable, or have their own specific parameters.
However, some parameters or custom data may not work with the message passing services, even if they are fine on each chain.
In particular, OFT, NTT, and CCT messages may require normalized decimal values for ERC-20 tokens (i.e., at most some value like $6$).

\subsubsection{Fees \& Performance.} 
Cross-chain token transfers incur costs across multiple dimensions: gas fees on source and destination chains, messaging and verification fees paid to intermediary infrastructure, and potential protocol-level fees. 
All five standards require gas payments on both source and destination chains for burn/lock and mint/unlock operations, though they differ in how destination gas is funded. 
OFT, CCT, NTT, and Connext xERC20 implementations typically allow users to prepay destination gas in the source transaction via relayer services, improving user experience by enabling single-transaction cross-chain transfers~\cite{layerzero-fees, ccip-billing, wormhole-ntt, connext-fees}.

The standards diverge significantly in messaging and protocol fees. 
The xERC20 standard imposes no protocol-level fee; costs depend entirely on the chosen bridge implementation. 
The Connext implementation of xERC20 charges a 0.05\% fee for accelerated liquidity through its fast-path router service. 
OFT charges usage-based oracle and relayer fees that scale with message payload size and destination gas requirements rather than transfer value~\cite{layerzero-fees}. 
Notably, OFT includes a governance-controlled fee switch that could enable protocol fees via governance decisions, though it remains disabled as of this writing~\cite{layerzero-oft-reference}.

NTT similarly has no protocol fee. Wormhole's 19-node guardian network provides message attestations without per-message charges. Users may optionally pay relayer fees to cover destination gas for automatic delivery, but this fee is entirely gas-based with no protocol markup~\cite{blockworks-ntt}. 
CCT actively charges a network fee under Chainlink’s CCIP billing model, in Chainlink's LINK token. 
The network fee structure depends on the token pool mechanism: lock-and-unlock transfers incur percentage-based fees (0.05\% for non-LINK payments, 0.063\% for LINK), while burn-and-mint, lock-and-mint, and burn-and-unlock transfers incur fixed USD-denominated fees, with LINK payments receiving a discount on these fixed fees. 
Fixed fees also vary by lane: non-Ethereum lanes incur lower fees than transfers originating from or destined for Ethereum~\cite{ccip-billing}.

SuperchainERC20, operating within the OP Stack’s native L2-to-L2 messaging infrastructure, incurs only standard L2 gas costs with no additional messaging or protocol fees~\cite{optimism-superchain}. 

Performance characteristics follow directly from each standard's architecture design. 
Standards that rely on third-party message relayers, like OFT, NTT, and CCT introduce additional latency due to multi-party attestation and cross-chain finality requirements. 
In contrast, SuperchainERC20 is designed to achieve one-block latency within the Superchain by relying on shared execution and settlement assumptions across OP Stack chains.

\subsubsection{Security Model.}
All standards except SuperchainERC20 support explicit rate limiting as a mitigation mechanism against cross-chain supply shocks or attacks.
However, only xERC20 and OFT lets issuers decide which relayers are trustworthy within a message passing setting.
No standard allows the quorum threshold of a bridge to be overridden, though xERC20's ability to use multiple bridges provides a way to bypass this (e.g., setting up one's own bridge).
Moreover while most contracts can be paused (because they are issuer-controlled), in the case of CCT and Superchain, the messages sent to a paused contract may persist until it is unpaused.
This means extra care needs to be used in such a scenario.

\subsection{Evaluation}
\label{sec:evaluation}

To answer questions about adoption and prevalence, we collected data for tokens using these standards according to the following methodology.
All data was collected on 15 December 2025.

\subsubsection{Data Collection and Methodology}
\label{sec:methodology}

Data collection methodology differed slightly for each standard, and we describe each approach below.

\begin{itemize}
    \item \emph{xERC20.} As there is no central protocol for xERC20, deployments were identified using the publicly maintained Connext cross-chain token registry~\cite{connext_chaindata}. 
    Since this standard is proposed by Connext, Connexts's bridge is supposed to maintain the highest number of xERC20 tokens and this dataset reflects issuer-declared configurations captured in registry metadata. 
    \item \emph{OFT.} Deployment counts and network coverage for OFT tokens were obtained using LayerZero's public metadata API. 
    Specifically, we queried the \texttt{/v1/metadata/experiment/ofts/list} endpoint of the LayerZero API~\cite{layerzero_oft_api}.
    \item \emph{NTT.} Native Token Transfer (NTT) data was collected using WormholeScan's public API. 
    Token lists were retrieved from the \texttt{/api/v1/native\--token-transfer/token-list} endpoint~\cite{wormholescan_api}.
    \item \emph{CCT.} CCT deployment data was obtained from Chainlink’s CCIP token directory API. 
    We queried the \texttt{/api/ccip/v1/tokens} endpoint with the \texttt{environment=mainnet} parameter~\cite{chainlink_ccip_api}.
    \item \emph{SuperchainERC20.} The cross-chain functionality of the SuperchainERC20 standard is not in production yet, so we did not have any data to use in our analysis regarding tokens that use it.
\end{itemize}

Raw datasets obtained were normalized by token symbol and identifier. Where standards supported multiple contract instances or variants of the same token, entries were de-duplicated at the token level. 
Cross-standard overlap counts therefore represent tokens with deployments across multiple interoperability standards, rather than individual contract instances.

\subsubsection{Adoption}
\label{sec:adoption}

Table~\ref{tab:token-counts} summarizes observed deployment characteristics across interoperability standards.

\begin{table}[tb]
\centering
\caption{Token counts from official sources and API extraction, along with total value transferred through their official cross-chain protocols. No value is listed for xERC20 as it uses independent bridges to transfer value despite the standard being championed or developed by Connext. As Superchain cross-chain functionality is not live, we only list the number of networks that would benefit from its implementation at the time of writing.}
\label{tab:token-counts}
\small
\renewcommand{\arraystretch}{1.25}
\setlength{\tabcolsep}{10pt}
\begin{tabular}{l c c c} 
\toprule 
\textbf{Standard} & ~\textbf{Token} & \textbf{Network} & \textbf{Total Value Transferred} \\ 
 & ~\textbf{Count} & \textbf{Count} & \textbf{(USD)} \\ 

\midrule
\rowcolor{gray!20} xERC20          & 14  & 8 & - \\
OFT   & 356 & 75 & \$144,236,346,809 \\
\rowcolor{gray!20} NTT    & 121 & 55 & \$3,852,445,453 \\
CCT  & 214 & 56  & \$11,257,377,248 \\ 
\rowcolor{gray!20} SuperchainERC20 & - & 34  &  - \\ 
\bottomrule
\end{tabular}
\end{table}

OFT exhibits the largest observed deployment footprint, with 356 tokens identified across 75 networks.
NTT and CCT show less deployment profiles.
There were 121 NTT-based tokens across 50 networks and 214 CCT tokens across 56 networks.
xERC20 lacks a canonical registry, limiting systematic measurement of adoption.
Transfer volume further differentiates deployment maturity, which is also shown in the table. 
OFT accounts for the largest observed cumulative transfer volume, followed by CCT and NTT. Higher transfer volumes likely correlate with earlier adoption, broader chain coverage, and integration into high-throughput applications.

Cross-standard adoption remains limited. 
Table~\ref{tab:overlap-matrix-1} shows overall token overlap across the four interoperability standards. 
Only 31 tokens appear in two or more standards, with just four tokens, DEGEN, PUFETH, USDT, and wstETH, appearing in three. 
No token currently implements all four standards simultaneously.
Table~\ref{tab:overlap-matrix} also shows pairwise token overlap across the four interoperability standards, showing CCT and LayerZero have the highest number of shared tokens.

\begin{table*}[bt]
\caption{Cross-standard token overlap across interoperability standards.}
\label{tab:overlap-matrix-1}
\centering
\renewcommand{\arraystretch}{1.2}
\begin{tabular}{lcc}
\toprule
\textbf{Metric} & \textbf{Count}\\
\midrule
Tokens appearing in exactly 1 standard & 649 \\
Tokens appearing in $\ge 2$ standards & 31 \\ 
Tokens appearing in 3 standards & 4 \\
Tokens appearing in all 4 standards & 0 \\
\bottomrule
\end{tabular}

\end{table*}

\begin{table}[bt]
\centering
\caption{Pairwise token overlap across interoperability standards.}
\label{tab:overlap-matrix}
\small
\renewcommand{\arraystretch}{1.25}
\setlength{\tabcolsep}{10pt}
\begin{tabular}{lrrrr}
\toprule
            & \textbf{OFT} & \textbf{NTT} & \textbf{CCT} & \textbf{xERC20} \\
\midrule
OFT         & 356 & 8   & 23  & 1 \\
NTT         &     & 121 & 6   & 1 \\
CCT         &     &     & 214 & 1 \\
xERC20      &     &     &     & 14 \\
\bottomrule
\end{tabular}
\end{table}

This limited overlap suggests that issuers typically commit to a single interoperability ecosystem, likely due to deployment complexity, governance overhead, and the absence of standardized mechanisms for coordinating minting authority, rate limits, and failure handling across protocols. 

\subsection{Discussion}\label{sec:discussion}

The standards explored in this work aim to solve the same problem but take different approaches.
Most rely on a specific protocol to pass messages between the systems, rather than providing a decentralized or protocol-agnostic standard.
Additionally, these standards do not do much to enable the efficient coordination of finance in a multi-chain setting.

Reliance on a centralized bridging protocol suggests that many of these solutions are instead to find users for their respective protocols, especially when fees are required, rather than solving the core problems of liquidity fragmentation.
While there is nothing inherently wrong with such behavior, open standards that are truly permissionless are much more interesting in the long term.
Systematic approaches to solving liquidity (and user) fragmentation, like the adoption of so-called based rollups (see, e.g.,~\cite{taiko-based}), are already underway.
However, that approach requires further research and fundamental changes to L2 architectures that will require a monumental amount of work; it also does not capture the scenario of bridging to other L1 networks.
Practical permissionless approaches are therefore still necessary, and the message passing and token standards they will need, should be defined and refined until they are used.

The prevalence of a number of competing token standards complicates the token landscape in an already complex multi-chain setting.
Adoption of standards by wallets, decentralized exchanges, and other user-focused (decentralized) applications is meant to reduce complexity, not increase it.
Instead, theses competing standards do not provide consistent chain support, universal functionality, or convenience.
Moreover, the governance of the standards themselves -- when tied to specific protocols -- is not necessarily accessible or aligned with ecosystems like Ethereum.
This does not mean these are not useful starting points, but only that a convalescence of required and desired functionality is necessary to support truly open standards with protocol agnostic message passing.
A simpler starting place may be to present a unifying wrapped for a given chain, that can take representations of tokens in all standards and mint a single token.
This would be useful not for the same token issued via multiple standards (it appears unlikely that would happen), but so that if a chain's applications wish to trade different multi-chain tokens, they don't need to write wrappers for each standard.

\section{Conclusion}
\label{sec:conclusion}

This paper presented a comparative analysis of emerging cross-chain token standards, examining how different architectural choices shape security, governance, and interoperability trade-offs. 
We analyzed xERC20, OFT, NTT, CCT, and SuperchainERC20 tokens in order to understand their differences.

Our analysis reinforces that choice of cross-chain token standard reflects a set of trade-offs between verification strength, flexibility, latency, and governance assumptions. 
As multi-chain deployments continue to expand, making these trade-offs explicit becomes increasingly important for issuers, integrators, and users. 
We expect future interoperability efforts to build on these foundations by further decoupling token semantics from transport mechanisms while improving composability across heterogeneous execution environments.

\subsection{Future Work}
\label{sec:future-work}

This study highlights several directions for future research in cross-chain token interoperability. 
First, while the analysis focuses primarily on Ethereum compatible standards, extending the comparison to ecosystems such as Cosmos and Polkadot would help determine whether the architectural trade-offs identified here generalize to environments built around different consensus and messaging assumptions. 
Such cross-ecosystem analysis could clarify which designs represent universal patterns or are artifacts of EVM-based systems.

Second, the comparison in this paper is largely structural. Complementary empirical evaluation of cross-chain token standards under realistic workloads would provide quantitative grounding for the observed trade-offs. 
Measurements of end-to-end latency, throughput under load, and cost variability across standards could help characterize how verification depth, rate-limiting mechanisms, and queuing behavior affect performance in practice.

Third, the increasing complexity of protocol designs raises open questions about long-term security and governance. Systematic study of incident reports, failure modes, and audit findings across deployed cross-chain token systems could provide insight into how layered verification, monitoring networks, and administrative controls perform under adversarial conditions. Formal models of cross-chain token supply and relevant invariants may safeguard these systems in the future, as well as ease the adoption of these standards to other token types.

Finally, we emphasize the need to study the user experience and wallet integrations of these standards.
While a single winner is unlikely, if users are not able to benefit from the simplified architecture these standards aim to proliferate, then further work is necessary to overcome the initial problems of cross-chain token fragmentation, confusion, and complexity.

\bibliographystyle{splncs04}
\bibliography{Document}
\onecolumn

\appendix
\section{Implementation Details}\label{sec:impl-details}

In this section, we recall some details of the ERC-20 standard itself before providing illustrative excerpts highlighting the minimal function signatures and user-facing entry points relevant to standards considered in this work.
This is for reference in case the exposition around expected behavior is assisted by actual Solidity function signatures; there are no insights or novel claims within this appendix.
Note that these excerpts are not complete contract interfaces.

\subsection{ERC-20 Tokens}\label{sec:erc20tokens}

Recall that the ERC-20 token standard does not define any minting or burning functionality~\cite{ERC-20}. 
However, the widely used OpenZeppelin implementation\footnote{\url{https://docs.openzeppelin.com/contracts/5.x/erc20}} extends ERC-20 with internal \texttt{\_mint()} and \texttt{\_burn()} functions. 
These functions are not part of the ERC-20 specification itself; they are implementation details that derived contracts may customize if they require a supply mechanism. 
In a single-chain ERC-20 deployment, these internal functions are used solely to adjust the total supply on that chain and are typically restricted to privileged roles. 
\subsection{xERC20}\label{subsec:xerc20-impl}

xERC20 tokens are required to implement the following interface, as well as the interface for ERC-20 tokens:

\begin{lstlisting}[style=xcodelight]
interface IXERC20 {
    event LockboxSet(address _lockbox)

    function mint(user, amount) external
    function burn(user, amount) external

    function setLimits(bridge, mintLimit, burnLimit) external

    function mintingMaxLimitOf(bridge) external
    function burningMaxLimitOf(bridge) external
    function mintingCurrentLimitOf(bridge) external
    function burningCurrentLimitOf(bridge) external

    function setLockbox(lockbox) external
}
\end{lstlisting}

Lockbox contracts must implement the following interface:

\begin{lstlisting}[style=xcodelight]
interface IXERC20Lockbox {
    deposit(amount) external
    withdraw(amount) external
}
\end{lstlisting}

\subsection{OFT}\label{subsec:oft-impl}

OFT tokens are instances of \texttt{OFCore} and ERC20 tokens, which therefore require them to implement the following interface.

\begin{lstlisting}[style=xcodelight]

abstract contract OAppCore{
    function oAppVersion() external view
    function endpoint() external view
    function peers(...) external view
    function setPeer(...) external
    function setDelegate(...) external
}

abstract contract OAppSender is OAppCore {
    function oAppVersion() public view virtual
    function _quote(...) internal view virtual
    
    function _lzSend(...) internal virtual

    function _payNative(...) internal virtual
    function _payLzToken(...) internal virtual
}

abstract contract OAppReceiver is OAppCore {
    function oAppVersion() public view virtual
    function isComposeMsgSender(...) public view virtual

    function allowInitializePath(...) public view virtual
    function nextNonce(...) public view virtual
    function lzReceive(...) public payable virtual
    function _lzReceive(...) internal virtual
}

// Generic LayerZero messaging logic (endpoints, message verification, etc)
abstract contract OApp is OAppSender, OAppReceiver {
}

abstract contract OFTCore is OApp{
    function oftVersion() external pure virtual
    function sharedDecimals() public view virtual
    function setMsgInspector(...) public virtual onlyOwner
    function quoteOFT(...)external view virtual
    function quoteSend(...) external view virtual
    function send(...) external payable virtual
    function _buildMsgAndOptions(...) internal view virtual
    function _lzReceive(...) internal virtual override
    function _lzReceiveSimulate(...) internal virtual override
    function isPeer(...) public view virtual override
    function _removeDust(...) internal view virtual
    function _toLD() internal view virtual
    function _toSD() internal view virtual
    function _debitView() internal view virtual
    function _debit() internal virtual
    function _credit() internal virtual
}

abstract OFT is OFTCore, ERC20 {
    function token() public view
    function approvalRequired() external pure virtual
    // Burn on source chain in _debit(...)
    function _debit(...) internal virtual override
    // Mint on destination chain in _credit(...)
    function _credit(...) internal virtual override
}

// Alternative path for existing ERC20 tokens
abstract OFTAdapter is OFTCore {
    // Uses safeTransferFrom(...) to lock on src
    function _debit(...) internal virtual override
    // Uses safeTransfer(...)   to release on dst
    function _credit(...) internal virtual override
}
\end{lstlisting}

For existing ERC-20 tokens that do not support minting and burning, LayerZero
provides the \texttt{OFTAdapter}. 
This contract also inherits \texttt{OFTCore} but replaces the burn-and-mint logic with lock-and-release semantics: \texttt{safeTransferFrom} is used to lock tokens inside the adapter on the source chain, and \texttt{safeTransfer} is used to release tokens on the destination chain. This allows legacy ERC -20 tokens to gain multi-chain transferability without modifying their original token logic. 

Therefore OFTs can be created one in of the following two ways:

\begin{lstlisting}[style=xcodelight]
// Native burn/mint omnichain ERC-20
MyOFT is OFT {}
// Adapter-based omnichain adapter for existing ERC-20
MyOFT is OFTAdapter{}
\end{lstlisting}

Every OFT contract is an \texttt{OApp}, meaning it inherits the standard OApp messaging interface and, critically, maintains an explicit registry of its peer contracts on all supported chains.
This registry is stored in the \texttt{peers} mapping:

\begin{lstlisting}[style=xcodelight]
mapping(uint32 eid => bytes32 peer) public peers;
\end{lstlisting}

The \texttt{peer} entry binds each endpoint ID (\texttt{eid}) to the expected contract on that chain.
During every incoming message, the OFT contract checks that \texttt{msg.sender} corresponds to the registered peer.

\subsection{NTT}\label{subsec:ntt-impl}

\begin{lstlisting}[style=xcodelight]
interface IManagerBase{
    function quoteDeliveryPrice(...) external view
    function setThreshold(...) external
    function setTransceiver(...) external
    function removeTransceiver(...) external
    function isMessageApproved(...) external view
    function isMessageExecuted(...) external view
    function nextMessageSequence() external view
    function upgrade(...) external
    function pause() external
    function getMode() external view
    function getThreshold() external view
    function transceiverAttestedToMessage(...) external view
    function messageAttestations(...) external view
    function token() external view
    function chainId() external view 
}

abstract contract ManagerBase is IManagerBase{}

interface INttManager is IManagerBase {
    event TransferSent(...)
    event TransferSent(...)
    event PeerUpdated(...)
    event TransferRedeemed(...)
    event OutboundTransferCancelled(...)
    function transfer() external payable
    function completeOutboundQueuedTransfer(...) external payable
    function cancelOutboundQueuedTransfer(...) external
    function completeInboundQueuedTransfer(...) external
    function attestationReceived(...) external
    function executeMsg(...) external
    function tokenDecimals() external view
    function getPeer(...) external view
    function setPeer(...) external
    function setInboundLimit(...) external
}

interface IRateLimiter {
    function getOutboundQueuedTransfer(...) external view
    function getCurrentInboundCapacity() external view
    function getInboundQueuedTransfer() external view
}

abstract contract RateLimiter is IRateLimiter{}

abstract NttManager is INttManager, RateLimiter, ManagerBase {}

// wormhole cross-chain messaging interface
// forwards NTT messages between chains
interface ITransceiver {
    function getNttManagerOwner() external view
    function getNttManagerToken() external view
    function getTransceiverType() external view
    function quoteDeliveryPrice(...) external view
    function sendMessage(...) external payable
    function upgrade(...) external
    function transferTransceiverOwnership(...) external
}

\end{lstlisting}

\begin{lstlisting}[style=xcodelight]
interface INttToken {
    function mint(address account, uint256 amount) external
    function setMinter(address newMinter) external
    function burn(uint256 amount) external
}
\end{lstlisting}

Rate limiting is also a post-deployment configuration step performed on each \texttt{NTTManager}. 
NTT enforces global transfer capacity through two complementary mechanisms: a chain-wide outbound limit and per-chain inbound limits. 
The outbound limit caps the total amount of tokens that may leave a given chain within a rate-limit window, preventing excessive drainage of local liquidity. 
In contrast, inbound limits are configured individually for each peer chain, allowing the token owner to control how much incoming liquidity is permitted from each remote chain during the same window. 
Both limits operate at the \texttt{NTTManager} level and serve as protective flow controls that shape cross-chain token movement and help mitigate liquidity shocks or abuse. When a transfer exceeds the configured outbound or inbound rate limits, the \texttt{NTTManager} does not revert. 
Instead, the transfer is placed into a time-based queue associated with the corresponding digest (for inbound messages) or sequence number (for outbound messages). 
Queued transfers may be executed later once the rate-limit window has elapsed, or canceled by the original sender in the outbound case, allowing locked funds to be reclaimed.

\texttt{Transceiver} contracts are the fundamental components responsible for transmitting and receiving NTT messages. 
Each \texttt{NTTManager} acts as a \texttt{TransceiverRegistry}, enabling the token owner to configure the required attestation threshold by registering multiple Transceiver contracts. 
Every Transceiver operates as an independent messaging path, and all attestations for a given cross-chain transfer are accumulated on-chain. 
A transfer is executed only once a Transceiver has received an attestations for the same message, ensuring that no single messaging channel can compromise correctness or liveness.

\subsection{CCT}\label{subsec:cct-impl}
Chainlink does not introduce a specific interface for CCT token standars, only introduces a set of requirements that any CCT token should implement: 
\begin{lstlisting}[style=xcodelight]
Minimal Requirements

Registration:
    owner()
    getCCIPAdmin() // Recommended for new tokens

Burn/Mint Mode:
    mint(address account, uint256 amount)
    burn(uint256 amount)
    burnFrom(address account, uint256 amount)
    decimals()
    balanceOf(address account)

Lock/Release Mode:
    decimals()
    balanceOf(address account)
\end{lstlisting}

\begin{lstlisting}[style=xcodelight]
abstract contract TokenPool{
  event LockedOrBurned(...)
  event ReleasedOrMinted(...)
  function isSupportedToken(...) public view virtual
  function getToken() public view
  function getRmnProxy() public view
  function getRouter() public view virtual
  function setRouter(...) public onlyOwner
  function supportsInterface(...) public pure virtual override
  function lockOrBurn() public virtual override
  function releaseOrMint(...) public virtual override
  function getTokenDecimals() public view virtual
  function getRemotePools(...) public view returns
  function isRemotePool(...) public view
  function getRemoteToken(...) public view
  function addRemotePool(...) external onlyOwner
  function removeRemotePool(...) external onlyOwner
  function isSupportedChain(...) public view
  function getSupportedChains() public view
  function applyChainUpdates() external virtual onlyOwner
  function setRateLimitAdmin(...) external onlyOwner
  function getRateLimitAdmin() external view
  function getCurrentOutboundRateLimiterState(...) external view
  function getCurrentInboundRateLimiterState(...) external view
  function setChainRateLimiterConfigs(...) external
  function setChainRateLimiterConfig(...) external
  function getAllowListEnabled() external view
  function getAllowList() external view
  function applyAllowListUpdates() external onlyOwner
  }
}
\end{lstlisting}

\subsection{Superchain ERC20}\label{subsec:superchain-impl}

\begin{lstlisting}[style=xcodelight]
interface IERC7802 is IERC165 {
    event CrosschainMint(...)
    event CrosschainBurn(...)

    function crosschainMint(...) external
    function crosschainBurn(...) external
}

abstract contract SuperchainERC20 is ERC20, IERC7802 {
    // Only the TOKEN_BRIDGE is authorized to call these functions.
    function crosschainMint(...) external
    function crosschainBurn(address _from, uint256 _amount) external
}
\end{lstlisting}

\begin{lstlisting}[style=xcodelight]
contract SuperchainTokenBridge {
    event SendERC20(...);
    event RelayERC20(...);

    function sendERC20(...) external;
    function relayERC20(...) external;
}
\end{lstlisting}

\begin{lstlisting}[style=xcodelight]
contract L2ToL2CrossDomainMessenger {
    event SentMessage()

    event RelayedMessage()

    function crossDomainMessageSender() external view
    function crossDomainMessageSource() external view

    function sendMessage() external payable
    function resendMessage() external payable
    function relayMessage() external

    function messageNonce() external view
}
\end{lstlisting}

Note that in practice, an offchain component may be necessary to complete cross-chain paths automatically.
Optimism has built the \texttt{Autorelayer}off-chain component that listens to the \texttt{SendERC20} events on the source chain and delivers the encoded message to the destination chain, where it triggers the relay path, but others may be possible.

\end{document}